\documentclass[a4paper,11pt]{article}

\usepackage{jheppub-mod}

\usepackage{amssymb,amsmath}
\usepackage{graphicx}

\newcommand{\sig}{\varsigma}
\newcommand{\eps}{\varepsilon}

\newcommand{\ph}{\varphi}
\newcommand{\kap}{\varkappa}
\newcommand{\Kappa}{{\mathrm{K}}}

\newcommand{\strtns}{{\bar{\mu}}}

\newcommand{\be}{\begin{equation}}
\newcommand{\ee}{\end{equation}}
\newcommand{\ba}{\begin{eqnarray}}
\newcommand{\ea}{\end{eqnarray}}

\newcommand{\tens}[1]{{\boldsymbol{#1}}}

\newcommand{\pa}{\partial}

\newcommand{\cv}[1]{{\tens{\partial}}_{#1}}

\newcommand{\sss}[1]{{\scriptscriptstyle #1}}

\title{Thermodynamics of two black holes}

\author[a]{Pavel Krtou\v{s}}
\author[b]{Andrei Zelnikov}

\affiliation[a]{Institute of Theoretical Physics, Faculty of Mathematics and Physics,\\
Charles University in Prague,
V~Hole\v{s}ovi\v{c}k\'ach 2, Prague, Czech Republic}

\affiliation[b]{Theoretical Physics Institute, Department of Physics,\\ University of Alberta,
4-181 CCIS, Edmonton, Alberta, Canada T6G 2E1}

\emailAdd{Pavel.Krtous@utf.mff.cuni.cz}
\emailAdd{zelnikov@ualberta.ca}

\abstract{
We study a system of two charged non-rotating black holes separated by a strut. Using the exact solution of the Einstein--Maxwell equations, which describes this system, we construct a consistent form of the first law of thermodynamics. We derive thermodynamic parameters related to the strut in an explicit form. The intensive thermodynamical quantity associated with the strut is its tension. We call the corresponding extensive quantity the thermodynamical length and we provide an explicit expression and interpretation for it.}


\arxivnumber{1909.13467}

\renewcommand{\beforetochook}{\bigskip}
\toccontinuoustrue

\begin{document}
\maketitle
\flushbottom


\section{Introduction}
\label{sc:intro}

\vspace{-1ex}

The discovery of a thermodynamic nature of black holes \cite{Bekenstein:1973ur,Bardeen:1973gs} connected together two different branches of physics and drastically changed our understanding of spacetime and its relationship to quantum field theory. It has lead to a prediction of a mechanism of black hole evaporation due to the Hawking radiation \cite{Hawking:1974sw} and has lead to new interesting puzzles as well (see \cite{Carlip:2014pma} for a review).

Typically, the black hole thermodynamics is formulated for a class of static (or stationary) spacetimes containing a black hole. These spacetimes are characterized by a set of parameters such as a mass of the black hole, charge, angular momentum, etc. We can understand these parameters, as well as any function of these parameters, as observables on the space of spacetimes---on the thermodynamical space of states.

Among these observables, the important role plays the total mass of the spacetime, which in the case of a simple black hole, is directly its mass. In more general situations (e.g., not asymptotically flat spacetimes, cosmological context, etc.) its definition can be more problematic. However, there exist several approaches how to identify the total mass, see e.g.~\cite{Anabalon:2018ydc}. The presence of a black hole (or, in general, of any horizon) in the spacetime allows to define the entropy of the black hole as a quarter of the horizon area, and the Hawking temperature as the surface gravity on the horizon divided by ${2\pi}$. Depending on the complexity of spacetimes in the consideration, the black hole (and the whole spacetime) is characterized by further quantities.

The thermodynamics then formulates relations among these parameters. The first law of the thermodynamics relates a variation of the total mass in terms of a selected set of independent observables, where this set typically contains the entropy ${S}$. The variation of the mass then contains the ``heat'' term ${TdS}$ with the Hawking temperature ${T}$. Such a law reflects a non-trivial dependence of the total mass on other thermodynamical observables.

In the simplest case of the charged rotating asymptotically flat black hole the first law relates the mass, interpreted as the total energy ${\mathcal{E}}$, with entropy ${S}$, charge ${Q}$ and angular momenta ${J}$ as
\begin{equation}\label{firslawstandard}
    d\mathcal{E} = T d S + \Phi d Q + \Omega dJ\,.
\end{equation}
It is not surprising that the change of energy ${\mathcal{E}}$ can be written in terms of independent variations of quantities characterizing the system. The non-triviality of the first law lies in the fact that the corresponding intensive quantities have a clear physical interpretation: ${T}$ is the Hawking temperature, ${\Phi}$ is the potential on the horizon, and ${\Omega}$ is the angular velocity.

This observation has been generalized in many directions. A notion of horizon thermodynamics was proposed  \cite{Padmanabhan:2003gd}. It was also generalized to all kinds of black hole horizons, null surfaces, and cosmological horizons \cite{Sheykhi:2014rka,Cai:2005ra,Cai:2008mh,Chakraborty:2015hna,Hansen:2016wdg}.

Recently, many works \cite{Kastor:2009wy,Parikh:2005qs,Kubiznak:2016qmn,Hansen:2016wdg,Appels:2017xoe} paid an attention to black holes with anti-de~Sitter asymptotic, i.e., to the situation with a negative cosmological constant. It was proposed \cite{Parikh:2005qs,Kastor:2009wy,Kubiznak:2016qmn} that the cosmological constant can be also understood as a thermodynamical quantity, namely, that it can be  interpreted as a pressure, ${P=-\frac{\Lambda}{8\pi}}$. The associated extensive variable is then called the thermodynamic volume ${V}$. However, it turns out that, on the thermodynamics side, the mass of the black hole in this situations corresponds  to the enthalpy ${\mathcal{H}}$ rather than to the energy ${\mathcal{E}}$. Namely, the first law can be written as (see, e.g., \cite{Cvetic:2010jb,Kubiznak:2016qmn})
\begin{equation}\label{firstlawenthalpy}
    d\mathcal{H} = T d S + \Phi d Q + \Omega dJ + V dP\,.
\end{equation}
It contains additional ``work'' term associated with the cosmological constant. Clearly, with the pressure--volume interpretation, this term has a form characteristic for the enthalpy in the standard thermodynamics.

One could be cautious about the interpretation of the cosmological constant as a pressure and about words as enthalpy in the black hole context. However, let us emphasize, that regardless this interpretation, the first law \eqref{firstlawenthalpy} is highly non-trivial. It states, that there exists a quantity ${P}$, which complements other standard quantities ${S}$, ${Q}$, and ${J}$ in such a way that a variation of the mass with respect to the entropy ${S}$ with other quantities fixed is given by the Hawking temperature ${T}$; a variation with respect to the charge ${Q}$ with other variables fixed gives the electric potential on the horizon; etc. If we supplemented the standard observables ${S}$, ${Q}$, ${J}$ by a different arbitrarily chosen observable, the variations would not be given by the standard quantities ${T}$, ${\Phi}$, ${\Omega}$. The existence of ${VdP}$ term follows from a non-trivial integrability condition. And the fact, that ${P}$ is up to a factor the cosmological constant, is even more satisfactory.

A similar discussion appears also in other contexts. People considered black holes in different backgrounds characterized by various extra parameters: besides the mentioned AdS black hole \cite{Kastor:2009wy} they studied, for example, Taub-NUT spacetimes \cite{Ballon:2019uha,Kubiznak:2019yiu}, magnetized black holes \cite{Gibbons:2013dna}, or C-metrics describing accelerated black holes in asymptotically flat or AdS spacetimes \cite{Appels:2016uha,Appels:2017xoe,Abbasvandi:2019vfz,Anabalon:2018ydc}.

In all these generalized situations one has to extend the first law by additional terms corresponding to new ingredients of the system. For example, for a black hole interacting with a cosmic strings one has to include terms ${\ell\, d\strtns}$ \cite{Appels:2016uha,Appels:2017xoe} corresponding to each piece of the string. Here ${\strtns}$ is the tension of the string and ${\ell}$ is the conjugate variable called the thermodynamic length \cite{Appels:2016uha,Appels:2017xoe}. Similarly to the discussion of the cosmological constant interpreted as a pressure, the existence of the first law with these additional ${\ell d\strtns}$ terms is non-trivial and it reflects the integrability property of the system.

All systems with cosmic strings discussed in this context contain strings piercing a black hole and stretching up to infinity \cite{Griffiths:2006tk,Krtous:2005ej}. Because the string is described by a conical singularity these spacetimes  are not exactly asymptotically flat or AdS and the black hole may not be considered as an isolated system. It causes various problems with a definition of the total mass and requires some kind of renormalization of the parameters of the infinite system.

Therefore, it would be interesting to study a sufficiently complicated, but isolated, system of black holes with a nicely behaving asymptotic. A natural candidate is a pair of black holes kept in the equilibrium by an additional interaction: for example, by the electromagnetic field, and/or by a material object between them.

The solutions of the Einstein equations describing a set of neutral black holes along an axis kept in equilibrium by a thin strut between them is known for long time \cite{Bondi1964,IsraelKhan:1964,Gibbons:1980,Gautreau2008,Costa:2000kf}. There were attempts to generalize it to the charged case perturbatively \cite{Bonnor:1981ag} or to the rotating case \cite{Bicak:1985lua}. Later these have been generalized to higher dimensions and/or quadratic gravity \cite{Herdeiro:2009vd,Herdeiro:2010aq}. Physical properties of a pair of identical black holes with opposite charges (black diholes) have been studied in detail by Emparan and Teo \cite{Emparan:2001bb,Emparan:1999au,Herdeiro:2009vd}, where they managed to find out a compact expression for the horizon area and the surface gravity of the components. Later this dihole solution was generalized to the case of arbitrary masses and electric charges of the black holes \cite{Alekseev:2007re,Manko:2007hi}. If one of the component is naked singularity instead of black hole, then an equilibrium configuration without a string (strut) has been found \cite{Alekseev:2007gt,Alekseev:2011nj}. The metric describing two charged black holes looks very complicated, nevertheless very simple formulae for the horizon areas and their surface gravities have been found \cite{Manko:2008gb}. Recently a series of papers appeared, where these results have been generalized to a pair of charged rotating black holes \mbox{\cite{Manko:2013iva,Cabrera-Munguia:2015fsa,Manko:2017avt,Cabrera-Munguia:2018omi,Manko:2018iyn,Alekseev:2019ykf}}.

In this paper we study the exact solution of the Einstein equations (without cosmological constant) which describes a pair of electrically charged not-rotating black holes localized at a finite distance. Equilibrium of this system is achieved by the electromagnetic field and by a strut between the black holes. The strut is described by a conical singularity with an excess angle which defines the tension of the strut. The system in the consideration does not contain any conical singularity on the semiaxes pointing from the black holes to infinity. The advantage of this setup is that the length of the strut is finite and the whole configuration forms an isolated system in an asymptotically flat spacetime. Thermodynamics of the compact isolated system in equilibrium is typically very robust because it does not involve infinite energies, and parameters of the solution are well defined by conserved asymptotic global charges.

The system is characterized by 5 independent parameters: by masses ${m}$, ${M}$ of the both black holes, by their charges ${q}$, ${Q}$, and by a separation parameter ${R}$. For this system we were able to formulate the first law in the form
\be\label{firstlawpreview}
d\mathcal{H}=t\,ds+T\,dS+\phi\,dq+\Phi\,dQ-\ell\,d\tau\,.
\ee
Here, $\mathcal{H}=M+m$ is the total mass of black holes, interpreted as a kind of enthalpy of the system. ${S}$ and ${s}$ are entropies of the black holes given by a quarter of horizon areas, $T$ and $t$ are the Hawking temperatures of black holes defined in a standard way through the surface gravity at horizons, and $\Phi$, $\phi$ are the electric potentials at the horizons. The parameter $\tau$ is the tension of the strut and $\ell$ is the thermodynamic length of the strut.

Our non-trivial observation is that the Einstein equations allow this form of the first law, with ${s}$, ${S}$, ${t}$, ${T}$, ${q}$, ${Q}$, ${\phi}$, ${\Phi}$ given by the standard definitions, and the last term is proportional to the variation of the strut tension. In the next sections we will derive this result, compute a simple form the thermodynamic length~$\ell$, and provide a lucid interpretation for it.

Similarly to the cases discussed above, this result reflects a non-trivial nature of the dependence of the total mass on the thermodynamic quantities. Moreover, it shows that the quantity naturally complementing ${s}$, ${S}$, ${q}$, and ${Q}$ is the tension of the strut ${\tau}$.

The plan of our work is as follows: First, in Section~\ref{sc:geometry}, we describe the system of two black holes and in Section~\ref{sc:physquant} we review its physical characteristics. \nopagebreak In Section~\ref{sc:firstlaw}, we derive the first law of thermodynamics for neutral and charged black holes. We interpret the thermodynamic length in Section~\ref{sc:StrutTerm}, and we shortly discuss the Smarr formula in Section~\ref{sc:Smarr}. Finally, in Section~\ref{sc:summary} we summarize and discuss our results.

\section{Double black-hole solution}
\label{sc:geometry}

An asymptotically flat static solution of Einstein-Maxwell equations, which
describes two nonextreme Reissner-Nordstr\"om black holes in equilibrium was obtained in an explicit and rather simple form
in \cite{Alekseev:2007gt,Alekseev:2007re,Alekseev:2011nj,Manko:2007hi,Manko:2008gb}.  Using cylindrical Weyl coordinates the
corresponding metric and the electromagnetic vector potential can be written as
\be\label{metric}
ds^2=-f\,dt^2+f^{-1}\left[h^2(d\rho^2+dz^2)+\rho^2\,d\varphi^2\right]
\ee
\be
A_t=-\Phi\,,\quad A_\rho=A_z=A_\varphi=0.
\ee
Here $f,h$ and $\Phi$ are the functions of the coordinates $\rho$ and $z$. The
Reissner-Nordstr\"om black holes with mass and charge parameters $M,Q$
and $m,q$, and separation parameter ${R}$, are assumed to be localized on a symmetry axis. Their horizons are represented by coordinate-singular rods placed on the axis at ${(z_H-\Sigma,z_H+\Sigma)}$ and ${(z_h-\sigma,z_h+\sigma)}$, respectively. The separation parameter ${R}$ defines the coordinate distance between centers of both rods,
\be
R=|z_H-z_h|
\ee

\begin{figure}[tbp]
\centering
\includegraphics[scale=0.5]{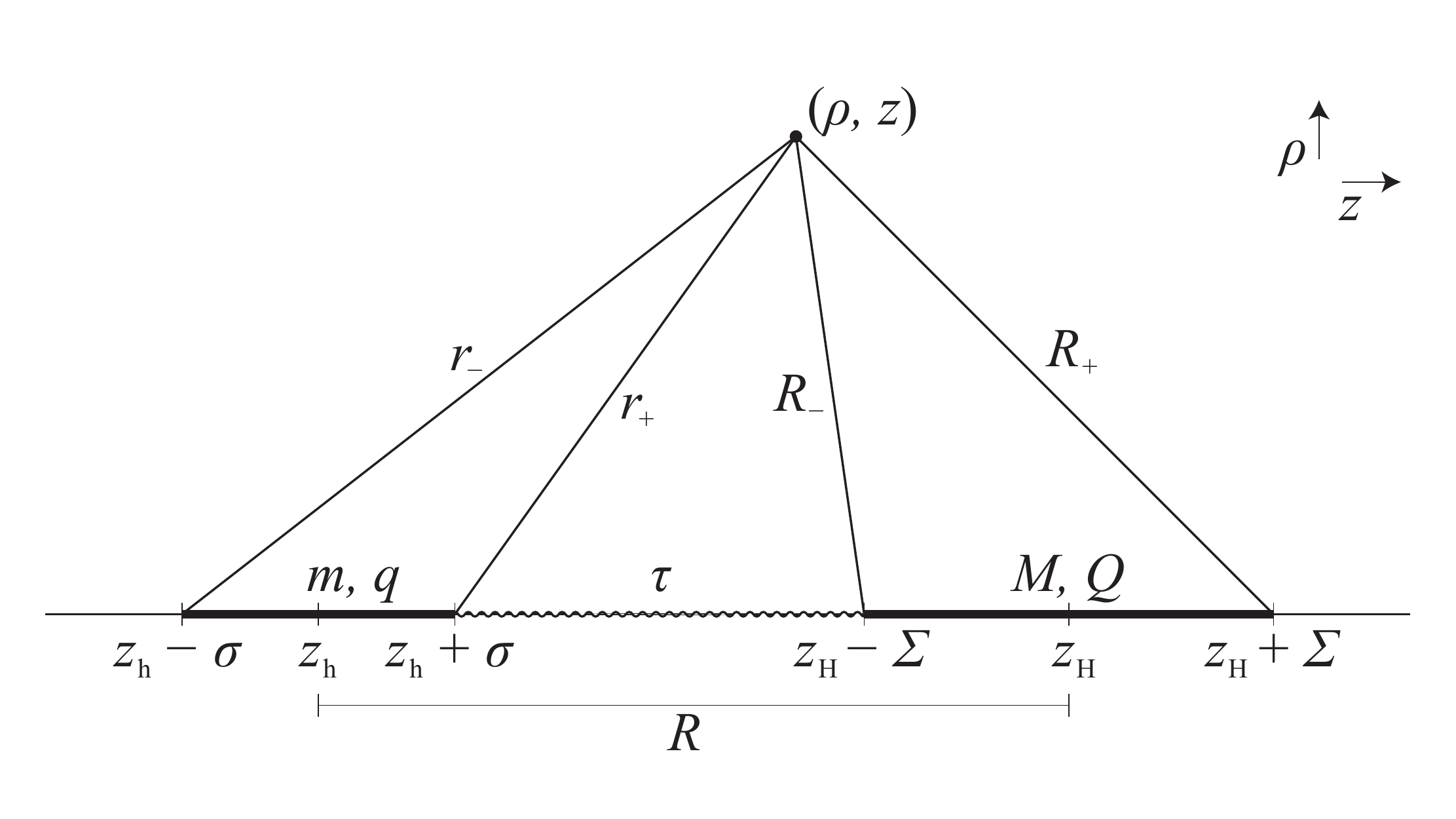}
  \caption{The horizons of black holes in Weyl coordinates ${\rho}$,~${z}$, (${t=\text{const}}$, ${\varphi=\text{const}}$) are depicted as coordinate-singular rods placed on the symmetry axis $\rho=0$. The centers of these rods correspond to the coordinates $z_h$ and $z_H$, the half-lengths of the rods are ${\sigma}$ and ${\Sigma}$. The quantities ${r_{\sss\pm}}$ and ${R_{\sss\pm}}$ are evaluated as the coordinate distance from the ends of the rods. They play a role in expressions for the metric functions.
\label{fig:2BHWeyl}}
\end{figure}

The functions $f$, $h$ and $\Phi$ get the simplest form
\cite{Alekseev:2007re,Alekseev:2011nj,Manko:2007hi,Manko:2008gb}, when expressed in terms of the ``coordinate distances'' from ends of the rods, cf.~Fig.~\ref{fig:2BHWeyl}:
\be\begin{aligned}
R_{\sss\pm}&=\sqrt{\rho^2+(z-z_H\mp\Sigma)^2}\,,\\
r_{\sss\pm}&=\sqrt{\rho^2+(z-z_h\mp\sigma)^2}\,.
\end{aligned}\ee
The half-lengths $\Sigma$ and $\sigma$ of the rods are given by
\be\label{sigma}
\Sigma^2=M^2-Q^2+2\mu Q\,,\quad
\sigma^2=m^2-q^2-2\mu q\,.
\ee
Here and below we use several constants:
\be\begin{gathered}\label{constsdef}
  \mu=\frac{mQ-Mq}{R+M+m}\;,\\
  \nu=R^2-\Sigma^2-\sigma^2+2\mu^2\,,\\
  \kap=M m -(Q-\mu)(q+\mu)\,,\\
  K_*=4\Sigma\sigma\bigl(R^2-(M-m)^2+(Q-q-2\mu)^2\bigr)\,.
\end{gathered}\ee

Using these variables one can write the functions $f$, $h$ in the form
\be\label{fh}
  f=\frac{\mathcal{A}^2-\mathcal{B}^2+\mathcal{C}^2}{(\mathcal{A}+\mathcal{B})^2}\,,\quad
  h^2=\frac{\mathcal{A}^2-\mathcal{B}^2+\mathcal{C}^2}
      {K_*^2 R_{\sss+} R_{\sss-} r_{\sss+} r_{\sss-}}\,.
\ee
The potential for the Maxwell field reads
\be
\Phi=\frac{\mathcal{C}}{\mathcal{A}+\mathcal{B} }\;.
\ee
Finally, functions $\mathcal{A}$, $\mathcal{B}$ and $\mathcal{C}$ are given by the
expressions
\be\begin{split}\label{Adef}
  \mathcal{A}={}&\Sigma\sigma\bigl[
    \nu(R_{\sss+}+R_{\sss-})(r_{\sss+}+r_{\sss-})
    +4\kap(R_{\sss+} R_{\sss-}+r_{\sss+} r_{\sss-})\bigr]\\
    &-(\mu^2\nu-2\kap^2)(R_{\sss+}-R_{\sss-})(r_{\sss+}-r_{\sss-})\,,
\end{split}\ee
\be\begin{split}\label{Bdef}
  \mathcal{B}=2\Sigma\sigma&\bigl[
    (\nu m+2\kap M)(R_{\sss+}+R_{\sss-})
    +(\nu M+2\kap m)(r_{\sss+}+r_{\sss-})\bigr]\\
    +2\sigma&\bigl[\nu\mu(Q-\mu)-2\kap(R M-\mu q-\mu^2)\bigr](R_{\sss+}-R_{\sss-})\\
    +2\Sigma&\bigl[\nu\mu(q+\mu)-2\kap(R m+\mu Q-\mu^2)\bigr](r_{\sss+}-r_{\sss-}),
\end{split}\raisetag{6.5ex}\ee
\be\begin{split}\label{Cdef}
  \mathcal{C}=2\Sigma\sigma&\bigl[
    \bigl(\nu(q+\mu)+2\kap(Q-\mu)\bigr)(R_{\sss+}+R_{\sss-})\\
    &+\bigl(\nu(Q-\mu)+2\kap(q+\mu)\bigr)(r_{\sss+}+r_{\sss-})\bigr]\\
    +2\sigma&\bigl[\nu\mu M +2\kap(\mu m-RQ+\mu R)\bigr](R_{\sss+}-R_{\sss-}) \\
    +2\Sigma&\bigl[\nu\mu m +2\kap(\mu M+Rq+\mu R)\bigr](r_{\sss+}-r_{\sss-})\,.
\end{split}\ee

The solution describes two black holes when quantities ${\sigma}$ and ${\Sigma}$ are real and satisfy the condition
\begin{equation}\label{BHcond}
    R>\Sigma+\sigma\,.
\end{equation}
The equality would correspond to the limit of touching black holes. Negative ${\sigma^2}$ or ${\Sigma^2}$ corresponds to the presence of a naked singularity instead of the black hole \cite{Alekseev:2007gt}.

\section{Physical quantities}
\label{sc:physquant}

The described solution has been thoroughly analyzed in \cite{Alekseev:2007gt,Alekseev:2007re,Alekseev:2011nj,Cabrera-Munguia:2018omi,Manko:2007hi,Manko:2008gb}. The most of physically interesting quantities has been calculated and we just list them here.

The total mass of the system is
\begin{equation}\label{energy}
    \mathcal{H} = M+m\;.
\end{equation}
The areas of horizons of both black holes are
\be\label{areas}
\begin{aligned}
    A&=4\pi\frac{\bigl((R+M+m)(M+\Sigma)-Q(Q+q)\bigr)^2}{(R+\Sigma)^2-\sigma^2}\,,
    \\
    a&=4\pi\frac{\bigl((R+M+m)(m+\sigma)-q(Q+q)\bigr)^2}{(R+\sigma)^2-\Sigma^2}\,,
\end{aligned}
\ee
the surface gravities are
\be\label{surfgrs}
\begin{aligned}
    \Kappa&=\frac{\Sigma\,\bigl((R+\Sigma)^2-\sigma^2\bigr)}
       {\bigl((R+M+m)(M+\Sigma)-Q(Q+q)\bigr)^2}\,,
    \\
    \kappa&=\frac{\sigma\,\bigl((R+\sigma)^2-\Sigma^2\bigr)}
       {\bigl((R+M+m)(m+\sigma)-q(Q+q)\bigr)^2}\,,
\end{aligned}
\ee
and the electric potentials on the horizons are
\be\label{potentials}
    \Phi=\frac{Q-2\mu}{M+\Sigma}\,,
    \quad
    \phi=\frac{q+2\mu}{m+\sigma}\,.
\ee

The total charges of each black hole are ${Q}$ and ${q}$, respectively. It is not a simple task to identify a mass of each black hole separately, since one cannot avoid a non-linear nature of the mutual interaction. But one can observe a remarkable property that both parameters ${M}$ and ${m}$ satisfy the Smarr relations in the form
\begin{equation}\label{sepSmarr}
    M = 2 T S + \Phi Q\,,
\quad
    m = 2 t s + \phi q\,,
\end{equation}
where entropies ${S}$, ${s}$ and temperatures ${T}$, ${t}$ of both holes are defined in the standard way
\begin{gather}
    \label{entropydef}
    S = \frac{A}{4}\,,
    \quad
    s = \frac{a}{4}\,,\\
    \label{tempdef}
    T = \frac{\Kappa}{2\pi}\,,
    \quad
    t = \frac{\kappa}{2\pi}\,.
\end{gather}
Therefore, we call ${M}$ and ${m}$ masses of the black holes.

Both black holes interact besides the gravitational and electromagnetic interaction also through a strut localized on the axis between them. It can be shown that the axis between black holes is not smooth but contains a conical singularity. Such a singularity represents a thin physical source with an internal energy and a tension. These can be related to the conical defect on the axis \cite{Israel:1976vc,Alekseev:2007re,Manko:2007hi}. When the angle ${\Delta\ph}$ around the axis is smaller than the full angle ${\Delta\ph = 2\pi-\delta}$, with $\delta>0$,  the object on the axis is called the cosmic string. It has a positive linear energy density ${\eps}$ and a tension ${\strtns}$ stretching the string (a negative linear pressure), which are related to the angular deficit ${\delta}$ as ${\strtns=\eps=\frac{\delta}{8\pi}}>0$. If the angle around the axis is bigger than $2\pi$, then $\delta<0$, and  the object represents the strut \cite{Israel:1976vc}. The strut has a negative energy density ${\eps}$ and a positive linear pressure $\tau$, which is called also the tension of the strut. These are related to the angular excess $-{\delta}$ as ${\tau=-\eps=-\frac{\delta}{8\pi}}>0$. Intuitively, because of the equality between linear energy density and tension, the effective gravitational masses of the string or the strut vanish. Its influence on a surrounding spacetime is also special: it effectively causes only the conical defect on the axis.

The discussed system contains the strut between the black holes with tension \cite{Manko:2007hi}
\begin{equation}\label{tension}
    \tau =\frac{\kap}{\nu-2\kap}=\frac{Mm-(Q-\mu)(q+\mu)}{R^2-(M+m)^2+(Q+q)^2} \,.
\end{equation}

\subsection*{Neutral black holes}

For uncharged black holes the quantities described above reduce to
\be\label{entropyNBH}
    S=4\pi M^2 \frac{R{+}M{+}m}{R{+}M{-}m}\,,
    \quad
    s=4\pi m^2 \frac{R{+}M{+}m}{R{-}M{+}m}\,,
\ee
\be\label{tempNBH}
    T=\frac{1}{8\pi M} \frac{R{+}M{-}m}{R{+}M{+}m}\,,
    \quad
    t=\frac{1}{8\pi m} \frac{R{-}M{+}m}{R{+}M{+}m}\,,
\ee
\be
    Q=q=0\,,\quad \Phi=\phi=0\;,
\ee
and
\be\label{tensionNBH}
\tau=\frac{M m}{R^2-(M+m)^2}.
\ee

\section{First law}
\label{sc:firstlaw}

Most of thermodynamic quantities describing this system have been known for a long time.  The thermodynamics has been discussed in the context of Majumdar-Papapetrou solutions describing special cases of extremal black holes, when the gravity and electromagnetism are in equilibrium and there are no strings or struts between them. Thermodynamics of a pair of black holes with equal masses and charges has been studied in the limit of large separation distance \cite{Costa:2000kf}. In higher dimensional context and for alternative gravity theories, it has been studied in \cite{Herdeiro:2009vd,Herdeiro:2010aq}. But we are not aware of an exact formulation of the first law for a simple system of two black holes with arbitrary  masses and charges.

Analogous situations with struts or cosmic strings have been recently discussed for the case of accelerated black hole in anti-de~Sitter spacetime \cite{Appels:2016uha,Appels:2017xoe,Abbasvandi:2019vfz,Anabalon:2018ydc}. It has been argued that the existence of the string (or similarly of the strut) implies an additional term in the first law. Such a term can be tentatively interpreted as a ``work'' term ${\ell\,d\strtns}$ where ${\strtns}$ is the tension of the string and ${\ell}$ is the thermodynamic length of the string. However, for the C-metric, the string is infinite and ${\ell}$ must be understood as a kind of renormalized length.

Following this motivation, we expect that the first law of the black hole thermodynamics for the double black hole system has the form
\be\label{firstlaw}
d\mathcal{H}=t\,ds+T\,dS+\phi\,dq+\Phi\,dQ-\ell\,d\tau\,.
\ee

To justify this assumption we have to express the total mass ${\mathcal{H}}$ (or at least its differential) in terms of thermodynamic quantities ${s}$, ${S}$, ${q}$, ${Q}$, and ${\tau}$. All these quantities are related to original parameters of the solutions ${m}$, ${M}$, ${q}$, ${Q}$, and ${R}$ by formulae \eqref{areas}, \eqref{entropydef}, and \eqref{tension}. So, it is enough to take differentials of these quantities, invert them and express ${dm}$, ${dM}$, ${dq}$, ${dQ}$, ${dR}$ in terms of ${ds}$, ${dS}$, ${dq}$, ${dQ}$, and ${d\tau}$. Substituting to ${d\mathcal{H}=dm+dM}$ then gives the first law. As a bonus, one derives the coefficients in front of the differentials ${ds}$, ${dS}$, ${dq}$, ${dQ}$, and ${d\tau}$ and identifies thus independently the temperatures, potentials, and the thermodynamical length.

Unfortunately, in the case of two charged black holes the expressions are too complicated to proceed exactly along this line. However, for neutral black holes, the described procedure can be followed exactly.

\subsection*{Neutral black holes}
Taking differentials of \eqref{entropyNBH} and \eqref{tensionNBH}, one gets
\begin{align}
  d s &= \frac{8\pi m}{(R{-}M{+}m)^2}
    \Bigl(\bigl[(R{+}m)^2-m M-M^2\bigr]\,dm + m(R{+}m)\,dM-m M\,dR\Bigr)\,,\notag\\
  d S &= \frac{8\pi M}{(R{+}M{-}m)^2}
    \Bigl(\bigl[(R{+}M)^2-m M-m^2\bigr]\,dM + M(R{+}M)\,dm-m M\,dR\Bigr)\,,\label{dsStauindmMR}\\
  d\tau &= \frac{1}{\bigl(R^2{-}(m{+}M)^2\bigr)^2}
    \Bigl(M\bigl(R^2{-}M^2{+}m^2\bigr)\,dm + m(R^2{+}M^2{-}m^2)\,dM - 2m M R \,dR\Bigr)\,.\notag
\end{align}
Solving for ${dm}$, ${dM}$ in terms of ${ds}$, ${dS}$, ${d\tau}$, we get
\begin{align}
  d m &= \frac{2R{-}m}{8\pi m (2R{-}M{-}m)}\frac{R{-}M{+}m}{R{+}M{+}m}\,ds
    -\frac{m}{8\pi M (2R{-}M{-}m)}\frac{R{+}M{-}m}{R{+}M{+}m}\,dS\notag\\
    &\mspace{260mu}-(R{+}M{-}2m)\frac{R{-}M{-}m}{2R{-}M{-}m}\frac{R^2{-}(M{+}m)^2}{R^2{-}(M{-}m)^2}\,d\tau\,,\notag
    \\[-0.75ex]
    \label{dmMRindsStau}
    \\[-0.75ex]
  d M &= \frac{2R{-}M}{8\pi M (2R{-}M{-}m)}\frac{R{+}M{-}m}{R{+}M{+}m}\,dS
    -\frac{M}{8\pi m (2R{-}M{-}m)}\frac{R{-}M{+}m}{R{+}M{+}m}\,ds\notag\\
    &\mspace{260mu}-(R{-}2M{+}m)\frac{R{-}M{-}m}{2R{-}M{-}m}\frac{R^2{-}(M{+}m)^2}{R^2{-}(M{-}m)^2}\,d\tau\,.\notag
\end{align}
Substituting to ${d\mathcal{H}=dm+dM}$, we obtain
\begin{equation}\label{firstlawNBH}
    d\mathcal{H}= t\,ds + T\, dS -\ell\,d\tau\,,
\end{equation}
where the temperatures ${t}$ and ${T}$ are indeed given by formulae \eqref{tempNBH} and the thermodynamical length ${\ell}$ of the strut is
\be\label{ellNBH}
  \ell= (R-M-m) \frac{ R^2-(M+m)^2}{R^2-(M-m)^2}\;.
\ee
We thus confirmed that the set of thermodynamical observables ${s}$, ${S}$, ${\tau}$ leads to the Hawking temperatures ${t}$, ${T}$ for the both black holes.

\subsection*{Charged black holes}

Calculations in the case of charged black hole are much more complicated. Fortunately, we know the expected form of the first law \eqref{firstlaw}, as well as temperatures \eqref{tempdef}, \eqref{surfgrs} and potentials \eqref{potentials}. The only missing quantity is the thermodynamical length ${\ell}$. To find it, we can compare ${d\mathcal{H}-t\,ds-T\,dS-\phi\,dq-\Phi\,dQ}$ with ${d\tau}$. The non-trivial fact proved in the calculation is that these differentials (expressed in ${dm}$, ${dM}$, ${dq}$, ${dQ}$, ${dR}$ basis) are indeed proportional. It justifies that the choice of the temperatures and potentials is correct.

However, the proportionality factor was a horribly complicated expression at the first glance. Nevertheless, we were able to extract this factor in a reasonable form by a tedious computation. Namely, we obtained that the thermodynamical length in the charged case~is
\be\label{elldef}
\begin{split}
\ell&=(R-\Sigma-\sigma)\frac{\nu-2\kap}{\nu+2\kap}\\
    &=(R-\Sigma-\sigma)\frac{R^2-(M+m)^2+(Q+q)^2}{R^2-(M-m)^2+(Q-q-2\mu)^2}\\
    &=(R-\Sigma-\sigma)\frac{(R^2-(M+m)^2+(Q+q)^2)^2}{(R^2-\sigma^2-\Sigma^2)^2-4\sigma^2\Sigma^2}\\
    &=\frac{(R^2-(M+m)^2+(Q+q)^2)^2}{(R+\sigma-\Sigma)(R-\sigma+\Sigma)(R+\sigma+\Sigma)}\,.
\end{split}
\ee

Independently of a derivation of ${\ell}$, with all expressions for the quantities in the first law in hand, one can check by a brute force that the first law is indeed satisfied---and verify thus the expression for ${\ell}$. This check involves again lengthy calculations and it has been done using algebraic manipulations on a computer.

Because the new term associated with the strut has form ${\ell\, d\tau}$, i.e., it contains the differential of the intensive observable ${\tau}$, the thermodynamical interpretation of the total mass is closer to an enthalpy than to an inner energy. We anticipated that already by choosing the letter ${\mathcal{H}}$ for the total mass. The situation is analogous to the case of the cosmological constant discussed in Section~\ref{sc:intro}, only in this case we deal with a linear pressure and length instead of a pressure and volume.

Note, however, that we obtained an opposite sign in front of the tension term than one would expect from the ordinary physics analogy. Indeed, the total mass ${\mathcal{H}}$, as well as both individual masses ${m}$ and ${M}$, decrease when enlarging the tension, with the entropies and charges fixed. In the uncharged case the variation of the individual masses can be seen in the formulae \eqref{dmMRindsStau}.

\section{The strut term}
\label{sc:StrutTerm}

\subsection*{Thermodynamic length}

The thermodynamic length \eqref{elldef} is always positive, provided that the condition \eqref{BHcond} is satisfied. It vanishes for $R=\Sigma+\sigma$, i.e., in the limit when the horizons touch.

However, ${\ell}$ is not the proper length between horizons of black holes. Such a length can be obtained by the integration of the length element along the axis and leads to elliptical integrals in parameters of the solutions.

It was actually Don Page, who in a discussion about meaning of the thermodynamic length suggested, that ${\ell}$ could be the length defined using the strut worldsheet area. This conjecture has turned out to be true. Later we found out that this interpretation of the thermodynamic potential conjugated to the strut tension was also proposed in \cite{Herdeiro:2009vd}.

\begin{figure}
\centering
\includegraphics[scale=0.5]{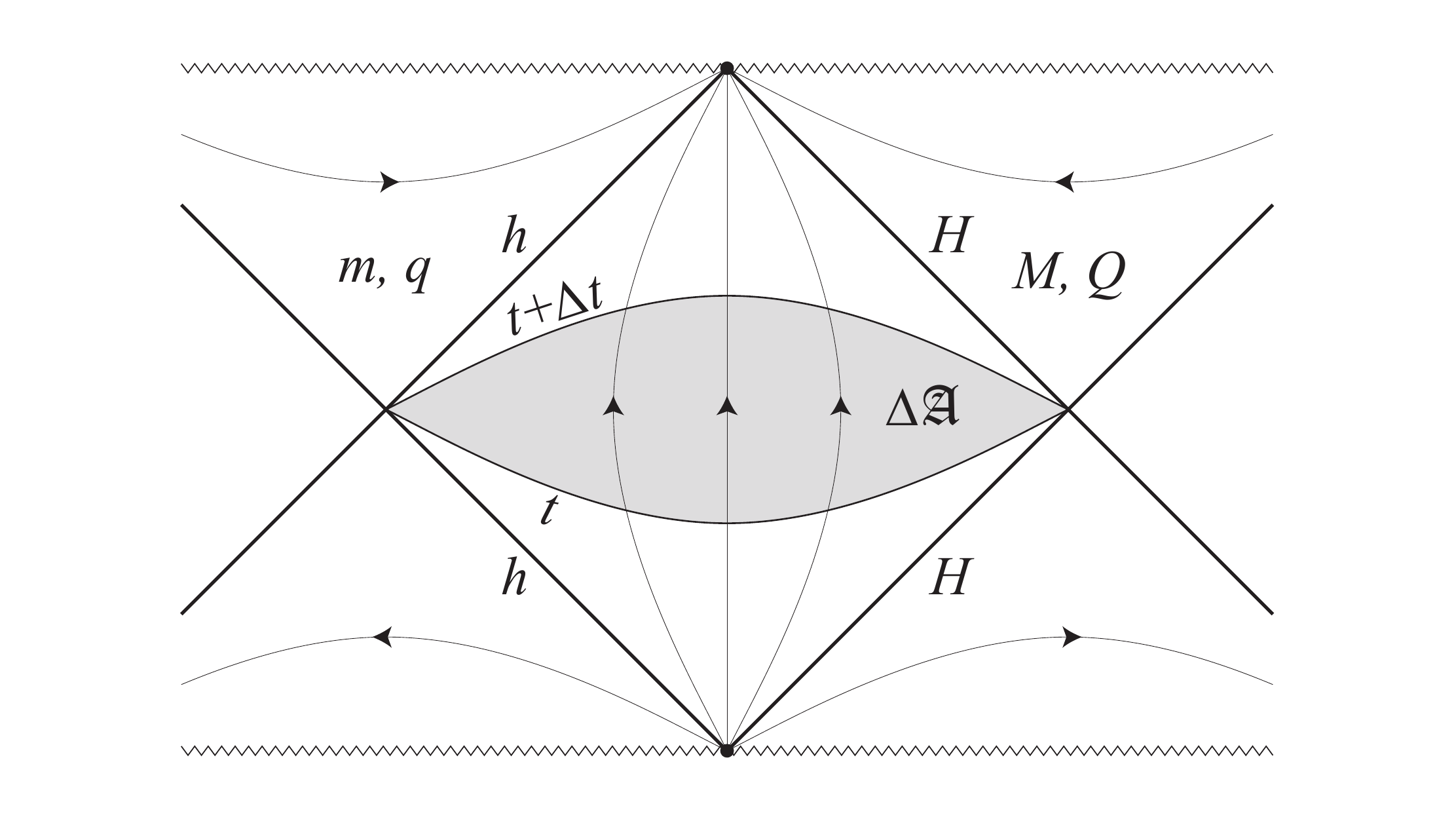}
  \caption{A schematic conformal diagram of the axis between black holes. The static domain between both holes is bounded by horizons ${h}$ and ${H}$. The orbits of the Killing vector ${\cv{t}}$ are indicated by lines with arrows. The worldsheet of the strut between two time lines ${t}$ and ${t+\Delta t}$ is in grey. It is spanned from the bifurcation neck of one horizon up to the bifurcation neck of the other horizon and it has area ${\Delta\mathfrak{A}}$. The zig-zag lines represent singularities inside black (and white) holes.
\label{fig:strutarea}}
\end{figure}

The area of the worldsheet of the strut between horizons during the static time interval ${\Delta t}$, cf.~Fig.~\ref{fig:strutarea}, is
\begin{equation}\label{strutarea}
    \Delta\mathfrak{A} = \int_{\text{strut}}d\mathfrak{A}
    = \int_t^{t+\Delta t}\int_{z_h+\sigma}^{z_H-\Sigma} h\, dz dt\,.
\end{equation}
It is straightforward (but lengthy) to show that the metric function ${h}$ on the axis between black holes is constant and given~by
\begin{equation}\label{honstrut}
    h\big|_{\text{strut}} = \frac{\nu-2\kap}{\nu+2\kap}\;.
\end{equation}
We thus obtain, that the thermodynamic length is given by the strut worldsheet area per unit of Killing time,
\begin{equation}\label{ellint}
    \ell = \frac{\Delta\mathfrak{A}}{\Delta t}\;.
\end{equation}

\subsection*{Relation of strut variables to the angular period}

The tension $\tau$ is directly given by the angle deficit ${\delta=-8\pi\tau}$ and, hence, by the period $\Delta\ph$ of the angle around the strut
\begin{equation}\label{strutangper}
\Delta\ph=2\pi-\delta=2\pi(1+4\tau)\,,
\end{equation}
Moreover, the thermodynamic length \eqref{elldef} has a structure ${\ell =L\, h\big|_{\text{strut}}}$ where
\begin{equation}\label{Ldef}
L = R-\Sigma-\sigma
\end{equation}
is the coordinate length of the strut between the horizons in the Weyl coordinates. The value of $h\big|_{\text{strut}}$ given by \eqref{honstrut} can be expressed in terms of the tension $\tau$ of the strut \eqref{tension}, and it is thus directly related to the angle period $\Delta\ph$,
\begin{equation}\label{struth}
h\big|_{\text{strut}}=\frac{1}{1+4\tau} = \frac{2\pi}{\Delta\ph}\,.
\end{equation}
Thus, the strut contribution $-\ell d\tau$ to the first law \eqref{firstlaw} can be represented as
\begin{equation}\label{elldtau}
-\ell\, d\tau = \frac{L \Delta\ph}{4}\, d\frac{1}{\Delta\ph}\,.
\end{equation}
In an analogy with temperature and area of the black hole we could introduce the \emph{infinitesimal area} $\alpha$ of an infinitesimal cylinder around the strut and the corresponding entropy~$\sig$
\begin{equation}\label{strutentr}
    \alpha = L\, \Delta\ph\,,\qquad
    \sig = \frac\alpha4 = \frac{L\, \Delta\ph}{4}\,,
\end{equation}
and a kind of a \emph{strut temperature} $\theta$ as an inverse angular period around the strut\footnote{Note that the \it{infinitesimal area} $\alpha$ and \it{temperature} $\theta$ of the strut have different dimensionality than those of the black hole horizon. It is related to the fact that we are using the dimensionless angular period ${\Delta\ph}$ instead of the period of Euclidean time used in the case of the black hole horizon.}
\begin{equation}\label{struttemp}
    \theta = \frac{1}{\Delta\ph}\,,
\end{equation}
which allow us to write the strut contribution to the first law as
\begin{equation}\label{adtht}
    -\ell\,d\tau = \sig\, d\theta\;.
\end{equation}

Although this form of the strut term looks very suggestive, we consider it just as a formal analogy with the heat terms of the black holes. We do not expect a thermal behavior of the strut similar to the Hawking radiation of the black hole. We will return to the interpretation of this term shortly in Section~\ref{sc:summary}.

Let us mention, that one could also rearrange the strut term as
\begin{equation}\label{LdlnDph}
  -\ell\, d\tau = - \frac{L}{4}\,d\ln\frac{\Delta\ph}{2\pi}\,.
\end{equation}
so the potential conjugated with a quarter ${\frac{L}{4}}$ of the coordinate length is ${\ln\frac{\Delta\ph}{2\pi}}$.

Of course, it is not surprising that one can choose different thermodynamical quantities to describe the contribution to the first law. The choice of conjugated pairs ${\{\ell,\tau\}}$, ${\{\sig,\theta\}}$, or ${\{\frac{L}{4},\ln\frac{\Delta\ph}{2\pi}\}}$ should be dictated by an application to particular physical processes and by a preferred parametrization of the system.

\section{Smarr formula}
\label{sc:Smarr}

The separate Smarr relations \eqref{sepSmarr} can be combined to the full Smarr formula \cite{Smarr:1972kt}
\begin{equation}\label{Smarr}
    \mathcal{H} = 2 t s + 2 T S + \phi q + \Phi Q \,.
\end{equation}
It does not contain any term related to the strut. It is not surprising since the terms in the Smarr formula depend on scaling properties of the involved variables. If the total mass scales as
\begin{equation}\label{scaling}
    c^p \mathcal{H}(X_1, X_2, \dots)
    = \mathcal{H}(c^{p_1} X_1, c^{p_2} X_2, \dots) \,,
\end{equation}
it implies the Euler formula
\begin{equation}\label{Euler}
    p \,\mathcal{H}(X_1, X_2, \dots)
    = p_1\frac{\pa\mathcal{H}}{\pa X_1} X_1+p_2\frac{\pa\mathcal{H}}{\pa X_1} X_2+\dots\,,
\end{equation}
where the partial derivatives ${\frac{\pa\mathcal{H}}{\pa X_i}}$ can be identified from the first law.
In our case, scaling length by ${c}$, the mass  ${\mathcal{H}=m+M}$ scales as ${c^1}$, entropies ${s}$, ${S}$ scale as ${c^2}$, charges ${q}$, ${Q}$ as ${c^1}$, and the tension ${\tau}$ as ${c^0}$, i.e., it remains unscaled. These scaling exponents lead exactly to the Smarr formula \eqref{Smarr} without any tension-related term.\footnote{%
Alternatively, the infinitesimal area $\alpha$ and the strut entropy $\sig$ scales as $c^1$; the angular period $\Delta\ph$ and the strut temperature $\theta$ remain unscaled. Therefore, we again obtain no strut contribution to the Smarr formula.}

\section{Summary}
\label{sc:summary}

We have been able to formulate the first law for the double charged black hole system in the form \eqref{firstlaw}. The missing ingredient was the term related to the strut between black holes. We have found the observable ${\ell}$ conjugated to the tension of the strut, which we call the thermodynamical length of the strut. It turns out that it is the strut worldsheet area per unit of Killing time. It is always positive, vanishes when the black holes touch and grows to infinity when the black holes are far away and the tension of the strut becomes negligible. It is proportional to the coordinate distance ${L=R-\sigma-\Sigma}$ between the horizons with the coefficient of proportionality given by the metric function ${h}$ evaluated on the strut. That is directly related to the angular period $\Delta\ph$ around the strut, cf.~\eqref{struth}.

We interpret the terms in the first law \eqref{firstlaw} as ``heat'' terms associated with horizon areas and ``work'' terms associated with electromagnetic field and with the strut. Since the strut term has form ${\ell d\tau}$, the total mass plays a role of the enthalpy in variable ${\tau}$.

We should notice that the first law contains two heat terms ${T dS}$ and ${t ds}$ associated with two black holes with different horizon temperatures, which suggests that the system is not in a thermal equilibrium. The investigated system is an exact solution of the classical Einstein-Maxwell equations. The spacetime around the black holes is static and thus is in a classical equilibrium. Moreover, the time Killing vector $\cv{t}$ is a generator of both horizons. However, all this indeed does not guarantee a thermodynamical equilibrium of the system, when quantum effects on this background are taken into account. The persistent discrepancy in temperatures is related to the fact that on the classical level (without taking into account the Hawking radiation) we do not have an agent which would restore the thermal equilibrium. It corresponds to a thermodynamical situation with two large reservoirs at different temperatures which do not exchange a heat because there is no agent which would transfer the heat, or, in the zeroth approximation, the agent is so weak that it can be neglected.

In the next approximation we could look for a heat flow between reservoirs, which, however, would be still weak compared to the reservoirs, and its backreaction on them could be neglected. It would correspond to a study of  test quantum fields on the fixed double black hole background.

The standard technique used in static situations with a horizon is to find the Hartle--Hawking state regular on the horizon. The temperature of such a state can be found using the Wick rotation to the Euclidean version of the spacetime. The regularity of the thermal state corresponds to a requirement of the absence of a conical singularity on the Euclidean horizon and it defines the corresponding temperature as the inverse period in the Euclidean time.

But this technique cannot be used in a generic case if more horizons with different temperatures are present. Some conical singularities in Euclidean version of the spacetime will necessarily survive --- the spacetime can be regularized by a choice the Euclidean time period only at one of the horizons. Remaining Euclidean conical singularities are not the result of the sources like strings and struts and are distinct from them \cite{Appels:2017xoe}.

However, as we have already said, we do not expect to find a quantum state in a thermal equilibrium with both black holes. Instead, we expect an existence of a state corresponding to a flow of particles from one black hole to another. To find such a state and its characterization by its behavior on both horizons would be interesting endeavor in the quantum field theory on curved background, which, however, goes beyond a scope of this paper.

Such a state would describe the system which is stationary, but not in a thermal equilibrium: two fixed black holes with a weak heat flow among them. This system would still be based on the exact solution of the Einstein equations studied in this paper. Only when investigating backreaction effects, one would have to take into account an influence of the quantum fields on the spacetime geometry.

Let us emphasize that the first law of black hole thermodynamics and the Smarr relation are derived using only the properties of the classical Einstein-Maxwell equations in Lorentzian signature. There are no curvature and stress-energy singularities on the horizons, except those related to the string/strut. We studied the properties of the exact solution of the Einstein-Maxwell equations in Lorentzian signature well defined both outside, inside, and on the horizon. Euclidean version of the solutions with multiple horizons may behave badly at some horizons, like in the Schwarzschild-de Sitter case, or may not exist at all, like for the Kerr solution. Therefore we prefer the discussion in the physical signature.

Different horizon temperatures is not an unknown situation in the black hole thermodynamics \cite{Cvetic:2018dqf}. In all situations where the static domain of the observer is bounded by two horizons, one associates temperatures and entropies with both horizon \cite{Kubiznak:2016qmn}. The simplest example is the Schwarzschild-de~Sitter black hole. Here the static domain is bounded by the black hole horizon and the cosmological horizon \cite{Dolan:2013ft}. The other example is the C-metrics, where besides a black hole horizon there is also an acceleration horizon. Even the simplest Reissner--Nordstr\"om and Kerr geometries have two horizons: inner and outer, each having it's own temperature. It's interesting that there are arguments \cite{Curir:1979,Frolov:2016gwl,Frolov:2017rjz,Cvetic:2018dqf} that in some cases one has to attribute negative temperature to the inner horizons. In our case horizon temperatures of both horizons are positive.

Thermodynamical interpretation of the black hole laws, for example relation of the horizon surface gravity and the black hole temperature, appears when one considers quantum fields on  the classical background geometries. These fields may be in equilibrium or non-equilibrium quantum state, depending on the choice of the vacua.

Despite the fact that the spacetime with two black holes does not describe the thermal equilibrium, it still makes sense to discuss the first law of the black hole thermodynamics. The black hole constituents of the system have thermal character and in the zeroth approximation (no interaction) we can speak about a balance of the energy. The first law provides us with the Gibbsian temperature: the slope of the curve of energy versus entropy \cite{Cvetic:2018dqf}. This law is `universal': it depends only on the gravitational theory under consideration and not the matter content \cite{Kubiznak:2016qmn} describing the thermal interaction between black holes. The first law \eqref{firstlaw} describes thermal properties of the black holes and the strut, which form together a self-consistent solution of the Einstein equations.

No other contributions than the zeroth order black hole and strut terms appear in the first law as long as we do not consider backreaction effect of quantum corrections. Of course it would be interesting to compute these corrections. In order to do this, one has to define a proper state first. Let's recall once more that in the case of spacetimes with multiple horizons the equilibrium Hartle--Hawking vacuum, which is closely related to the Euclidean version of the spacetime, does not exist, and one has to search for a state describing a stationary flow.

Another approach could be modifying the background spacetime by coupling to a vortex solution of a gauge field in such a way that the Euclidian version of the spacetime would be regular, e.g., along the lines of \cite{Dowker:1991qe}. It would be interesting to see if such an approach would modify the thermodynamical relations discussed here. But similarly to the study of the quantum field on the background of the two black holes, this goes beyond the analysis of the well defined classical solution we intended in this paper

Finally, let us return to the alternative form \eqref{adtht} of the strut contribution ${\sig\,d\theta}$ to the first law. The definitions of the strut entropy $\sig$ and of the strut temperature $\theta$ are motivated by an analogy with the Euclidian analysis used for the characterization of the Hawking--Hartle state. However, in the case of the strut, we are not dealing with a Euclidean conical singularity but with the real massive source entering the Einstein equations. Therefore, we do not suggest that the inverse angular period is a real temperature and that the strut has a thermal character.

One can speculate that this form of the strut contribution to the first law could be understood using the approach \cite{Herdeiro:2009vd}, when free energy of a system of black holes and a strut is obtained by computing a contribution of a boundary term of the Einstein action on the Euclidean manifold with conical singularities. In this approach the strut contribution to the black holes thermodynamics comes from a similar technique as the horizon contribution and, hence, it's not surprising that they have a similar form.

Although the interpretation of various terms in the first law for a pair of black holes may be open to a discussion, our observation that it can be formulated in the presented form is non-trivial. It reflects the integrability property of involved thermodynamic quantities. The derived thermodynamic length ${\ell}$ supplements other thermodynamical observables and makes thus the thermodynamical description complete.

\acknowledgments

We would like to thank Don Page for a valuable discussion and for the suggestion of the meaning of the thermodynamic length. We also appreciates numerous discussions with David Kubiz\v{n}\'ak, which have been a motivation for this work, and a discussion with Valeri Frolov about the system out of the thermal equilibrium. We thank David Kubiz\v{n}\'ak and Ji\v{r}\'i Bi\v{c}\'ak for pointing out some references which we missed. We also thank the referee for asking questions which lead to an improvement of our discussion of the investigated system.

P.K. was supported by Czech Science Foundation Grant 19-01850S. The work was done under the auspices of the Albert Einstein Center for Gravitation and Astrophysics, Czech Republic. P.K. also thanks the University of Alberta for hospitality. A.~Z. thanks the Natural Sciences and Engineering Research Council of Canada and the Killam Trust for financial support.

\bibliographystyle{JHEP}

\providecommand{\href}[2]{#2}\begingroup\raggedright\endgroup

\end{document}